\documentclass[prd,aps,twocolumn,nofootinbib]{revtex4}

\usepackage{amsmath}
\usepackage{graphicx}

\begin{document}

\title{Charge distribution of the gauge-mediation type Q ball}

\author{Shinta Kasuya$^a$, Masahiro Kawasaki$^{b,c}$}

\affiliation{
$^a$ Physics Division, Faculty of Science,
         Kanagawa University, Kanagawa 221-8686, Japan\\
$^b$ Institute for Cosmic Ray Research, the University of Tokyo, Chiba 277-8582, Japan\\
$^c$ Kavli Institute for the Physics and Mathematics of the Universe (WPI), 
        Todai Institutes for Advanced Study, the University of Tokyo, Chiba 277-8582, Japan}
     
\date{Octoberr 28, 2025}

\begin{abstract}
We numerically study the formation of the gauge-mediation type Q balls in the logarithmic square potential
on three-dimensional lattices. We obtain the broad charge distribution of the Q ball of this type for the first time, to the best of our knowledge.
The charge of the Q ball at the peak of the distribution is smaller than what we estimated as the average of the largest 
tens of the Q balls in the logarithmic potential for the same initial amplitude of the field at the onset of its oscillation. 
We also discuss some impacts of the broad distribution on cosmology and astrophysics. In the B ball (Q being the baryon number) 
case, the broad distribution would lead to the coexistence of both stable and unstable B balls. We find that stable B balls can 
account for the dark matter of the universe without affecting successful big bang nucleosynthesis by the decay of the unstable
B balls, but the baryon number of the universe cannot be explained by them. On the other hand, the large L balls (Q being the 
lepton number) would be the dark matter as well while avoiding the constraints on the X and/or gamma rays from the decay of 
the smaller L balls.
\end{abstract}

\maketitle

\section{Introduction}
The Q ball is the energy minimum configuration of the scalar fields for the fixed charge $Q$ \cite{Coleman:1985ki}.
The Q-ball solution exists for the flat potential, which is naturally realized in the supersymmetric 
theory \cite{Kusenko:1997zq,Kusenko:1997si,Enqvist:1997si}. Large Q balls can be a good candidate of the dark matter of the universe 
\cite{Kusenko:1997si,Kasuya:2001hg,Kasuya:2000sc}, may simultaneously provide baryon asymmetry of the universe
\cite{Kusenko:1997si,Kasuya:2001hg,Kasuya:2000sc}, or they are long lived and decay into lighter particles to affect 
cosmologically  or astrophysically \cite{Kasuya:2005ay,Kasuya:2007cy,Kasuya:2024ldq}.

Large Q balls can form through the Affleck-Dine mechanism \cite{Kusenko:1997si}. In Refs.~\cite{Kasuya:1999wu,Kasuya:2001hg},
we investigated the formation of the so-called gauge-mediation type Q balls using three-dimensional lattice simulations 
for the potential \cite{Kusenko:1997si}
\begin{equation}
V(\Phi) = m^4 \log\left(1+\frac{|\Phi|^2}{m^2}\right),
\label{log-pot}
\end{equation}
and estimated the charge of the formed Q balls by the largest ones as
\begin{equation}
Q=\beta\left(\frac{\phi_0}{m}\right)^4,
\label{Qphi1}
\end{equation}
where $\phi_0$ is the amplitude of the field at the onset of the oscillation and $\beta\simeq 6\times 10^{-4}$. 

Although we determined the formed Q-ball charge monochromatically, since those Q balls  seemed to dominate the 
energy density of the Q balls, a lot of Q balls with smaller charges are also produced in the simulations.
In Ref.~\cite{Hiramatsu:2010dx}, the authors studied the Q-ball formation in the gravity-mediation using lattice simulations,
and found a rather broad distribution of the charge. The distribution of the gauge-mediation type in the potential (\ref{log-pot}) 
was estimated in Ref.~\cite{Kasuya:2010vq}, but very roughly. Therefore, we should investigate the charge distribution of the
gauge-mediation type Q balls more thoroughly, and consider their cosmological and astrophysical consequences.

In this article, we numerically study the formation of the gauge-mediation type Q balls on three-dimensional lattices.
We adopt the logarithmic square potential, derived from the two-loop calculation in Ref.~\cite{deGouvea:1997afu},
instead of Eq.~(\ref{log-pot}). This may result in the fact that larger Q balls of this type could form, since the gauge-mediation
potential would dominate over the gravity-mediation potential up to a bit larger amplitude compered to the potential (\ref{log-pot}).

The consequence of the broad charge distribution is that both stable (or long-lived) and unstable (or short-lived) Q balls
exist. Thus one must take into account the influence of the decay of the unstable Q balls.

The structure of the article is as follows. In the next section, we briefly review the gauge-mediation type Q ball. In Sec.~III, 
we explain the set up of our simulations and show the results, including the charge distribution of the formed Q balls.
We exemplify some consequences of the rather broad distribution for cosmology and astrophysics in Sec.~IV. Section V is
devoted to our conclusions.

\section{Gauge-mediation type Q balls}
\begin{figure}[ht!]
\includegraphics[width=60mm]{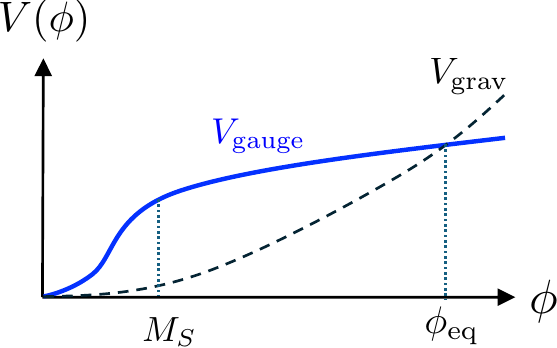} 
\caption{Sketch of the potential, where the gauge-mediation potential (\ref{gauge-pot}) is shown in blue solid line, 
while the gravity-mediation potential (\ref{grav-pot}) is denoted by the black dashed line.
\label{fig1}}
\end{figure}

The Q ball is the energy minimum configuration of the scalar fields for the fixed charge $Q$ \cite{Coleman:1985ki}.
In the minimal supersymmetric standard model, the scalar fields $\Phi$ is one of the flat directions, which are all classified in terms 
of gauge-invariant monomials \cite{Gherghetta:1995dv,Dine:1995kz}. 
In the gauge-mediated supersymmetry breaking scenario, the scalar potential is written as
\begin{equation}
V(\Phi )=V_{\rm gauge}(\Phi )+V_{\rm grav}(\Phi ),
\label{pot-all}
\end{equation}
where 
\begin{equation}
V_{\rm gauge}(\Phi)=\left\{ 
\begin{array}{ll}
m_\phi^2|\Phi|^2 & (|\Phi|\ll M_S), \\[2mm]
\displaystyle{M_F^4 \left(\log\frac{|\Phi|^2}{M_S^2}\right)^2} &  (|\Phi|\gg M_S),
\end{array}
\right.
\label{gauge-pot}
\end{equation}
is the gauge-mediation potential \cite{deGouvea:1997afu}, and $M_S$ is the messenger scale. 
$M_F$ is related to the $F$ component of a gauge-singlet chiral multiplet in the messenger sector, and its range is 
given by \cite{Kasuya:2015uka}
\begin{equation}
4\times 10^4~{\rm GeV} \lesssim M_F \lesssim 0.1 \left(m_{3/2}M_{\rm P}\right)^{1/2},
\label{MF}
\end{equation}
where $m_{3/2}$ is the gravitino mass and $M_{\rm P}=2.4\times 10^{18}$~GeV is the Planck mass. 
On the other hand, 
\begin{equation}
V_{\rm grav}(\Phi)= m_{3/2}^2|\Phi|^2,
\label{grav-pot}
\end{equation}
is the gravity-mediation potential. The sketch of the potential is shown in Fig.~\ref{fig1}.

The gauge-mediation type Q balls forms if the first term of the potential (\ref{pot-all}) dominates 
over the second one. This happens for the field values smaller than 
\begin{equation}
\phi_{\rm eq} \simeq \xi\frac{\sqrt{2}M_F^2}{m_{3/2}},
\label{phieq}
\end{equation}
where we define $\Phi=\frac{1}{\sqrt{2}}\phi e^{i\theta}$, and the factor $\xi$ is determined numerically
for the logarithmic square potential, approximately fitted as
\begin{eqnarray}
\xi & = & 44.5 +9.8\log_{10}\left(\frac{M_F}{10^6~{\rm GeV}}\right) \nonumber \\
& & -4.8\log_{10}\left(\frac{m_{3/2}}{\rm MeV}\right)
-4.9\log_{10}\left(\frac{M_S}{10^7~{\rm GeV}}\right).
\end{eqnarray}
In the following, we set $\xi=45$, for simplicity. The maximum amplitude of the field is then estimated as
\begin{equation}
\phi_{\rm max} \simeq 0.64 M_{\rm P} \simeq 1.6\times 10^{18}~{\rm GeV},
\end{equation}
where we insert the upper bound of (\ref{MF}) into Eq.(\ref{phieq}).

The properties of the gauge-mediation type Q ball are as follows: 
the mass and radius of this type of the Q ball are respectively given by
\begin{eqnarray}
\label{MQ}
M_Q & \simeq & \frac{4\sqrt{2}\pi}{3}\zeta M_F Q^{3/4}, \\
R_Q & \simeq & \frac{1}{\sqrt{2}}\zeta^{-1}M_F^{-1} Q^{1/4},
\label{RQ}
\end{eqnarray}
where $\zeta \simeq 2^{1/4}\sqrt{c_0/\pi}$ with $c_0\simeq 4.8\log(m_\phi/\sqrt{2}\omega_Q)+7.4$ 
\cite{Hisano:2001dr,Kasuya:2012mh,Kasai:2024diy}. In the following, we adopt $\zeta=5$, since we have
$m_\phi \simeq 10^4$~GeV and $\omega_Q \simeq 0.5$~MeV$-$1~GeV.
The rotation speed of the field inside of the Q ball reads
\begin{equation}
\omega_Q \simeq \sqrt{2}\pi\zeta M_F Q^{-1/4},
\label{omegaQ}
\end{equation}
equivalent to the mass per unit charge of the Q ball. The fact that $\omega_Q$ depends on $Q$ nontrivially 
is crucial for the criterion of the stability against the decay into lighter particles which have the same kind of the charge.

\section{Q-ball formation}
Let us study the formation of the gauge-mediation type Q balls. To this end, we numerically solve the equation
of the field $\Phi$ on the three-dimensional lattices in the potential 
\begin{equation}
V=4m^4\left[\log\left( 1+\frac{|\Phi|}{m}\right)\right]^2,
\end{equation}
which is equivalent to the potential (\ref{gauge-pot}) when the mass parameters are denoted in terms of $m$ as 
$m_\phi=2m$, $M_F=m$ and $M_S=m$, which we set for numerical feasibility. 
In the following, we show the cases of the box size $N=1000$, although we check that they are consistent with those cases with
a smaller box size of $N=512$.

In order to see the Q-ball formation, we solve the field equation, in which we decompose the complex scalar field $\Phi$ 
into its real and imaginary parts as $\Phi=\left( \phi_R+i\phi_I \right)/\sqrt{2}$, and normalize all the dimensionful 
parameters with respect to $m$ as $\varphi_\alpha= \phi_\alpha/m$ ($\alpha=R, I$), $\xi_i=mx_i$ ($i=1, 2, 3$), $\tau=mt$, $h=H/m$ and
$v=V/m^4$. Then the field equation can be written as
\begin{equation}
\varphi_\alpha''+3h\varphi_\alpha'-\frac{1}{a^2}\nabla_\xi^2\varphi_i+\frac{\partial v}{\partial \varphi_\alpha}=0 \quad (\alpha=R, I),
\end{equation}
where the prime denotes the derivative with respect to $\tau$.
Since the field $\Phi$ starts its oscillation (rotation) when the Hubble parameter becomes
$H_{\rm osc}=m_{\rm eff}(\phi_0)\equiv \sqrt{|V''(\phi_0)|}$, the initial time is estimated as 
$t_{\rm init}=2/(3H_{\rm osc})$, where the matter domination is assumed. We thus investigate the evolution of the scalar field
with initial conditions
\begin{eqnarray}
&& \varphi_R(\tau_{\rm init}) = \varphi_0 (1 + \Delta_1), \quad \varphi_R'(\tau_{\rm init}) = \Delta_2, \nonumber \\
&& \varphi_I(\tau_{\rm init}) = \Delta_3, \quad \varphi_I'(\tau_{\rm init}) = \varphi_0' (1 + \Delta_4), \nonumber \\
&& \tau_{\rm init}\equiv \frac{1}{3}\left(1+\frac{\varphi_0}{\sqrt{2}}\right)\left[\log\left(1+\frac{\varphi_0}{\sqrt{2}}\right)-1\right]^{-1/2},
\end{eqnarray}
where $\Delta$'s represent the fluctuations that originated from the quantum fluctuations of the field $\Phi$ during inflation and
the amplitudes are estimated as $O(10^{-7})$ compared with the corresponding homogeneous modes. We set the initial 
velocity of the $\varphi_I$ as
\begin{equation}
\varphi_0' = \left.\varphi\sqrt{\frac{v'}{\varphi}}\right|_{\varphi_0}=\left[ 4\sqrt{2}\varphi_0\frac{\log\left(1+\frac{\varphi_0}{\sqrt{2}}\right)}{1+\frac{\varphi_0}{\sqrt{2}}}\right]^{1/2},
\end{equation}
so that the orbit in the field space becomes circular if there is no cosmic expansion.

\begin{figure}[ht!]
\includegraphics[width=90mm]{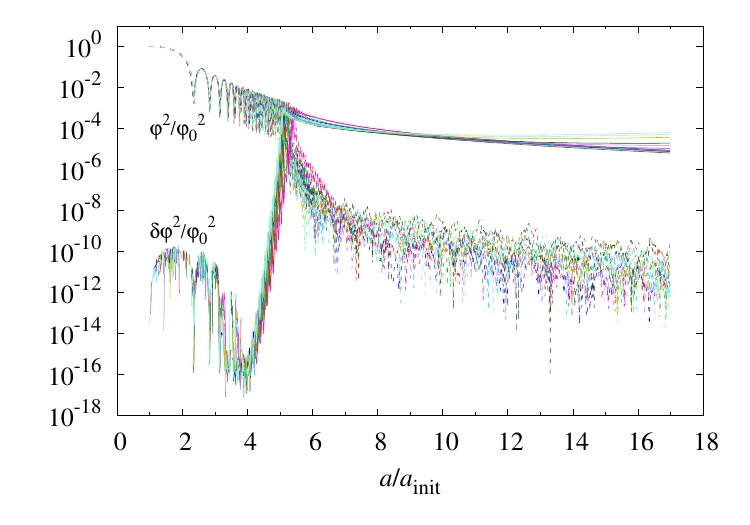} 
\caption{Evolution of the homogeneous modes $\varphi^2=\varphi_R^2 + \varphi_I^2$ and the fluctuations 
$\delta\varphi^2=\delta\varphi_R^2+\delta\varphi_I^2$ for various initial amplitudes.
\label{fig2}}
\end{figure}

We show the evolutions of the field $\Phi$ for the initial amplitudes of $\varphi_0=10^3$, $2\times 10^3$, $3\times 10^3$, 
$5\times 10^3$, $10^4$, $2\times 10^4$, $3\times 10^4$, and $5\times 10^4$ in Fig.~\ref{fig2}. Here, the dashed and solid lines
represent homogeneous modes and fluctuations, respectively. As can be seen, the evolutions of the field are almost identical 
in all the cases when they are normalized with respect to $\varphi_0$.

\begin{figure}[ht!]
\includegraphics[width=85mm]{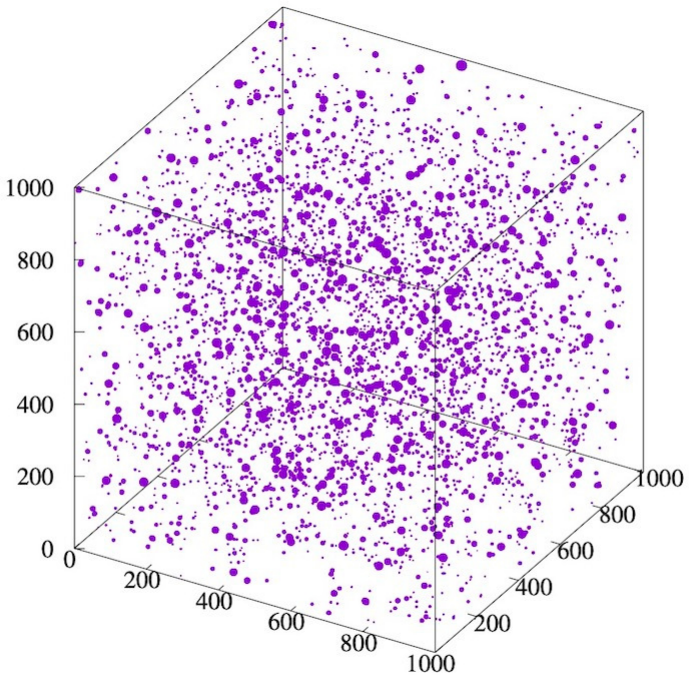} 
\caption{Formed Q balls in three-dimensional lattices with $N=1000$ and $\Delta\xi=0.5$ at $a/a_{\rm initi}\simeq 17$ 
for $\varphi_0=5\times 10^3$.
\label{fig3}}
\end{figure}

Homogenous condensates start to fragment into lumps at around $a/a_{\rm init}=5-6$ to form Q balls. We estimate the charges of
the formed Q balls at around $a/a_{\rm init}\simeq 17$ so that we avoid the possible initial excitation states of the Q balls. In each 
case, more than three thousand formed Q balls are identified in the simulation box, shown in Fig.~\ref{fig3} for 
$\varphi_0=5\times 10^3$ for example.

\begin{figure}[ht!]
\includegraphics[width=90mm]{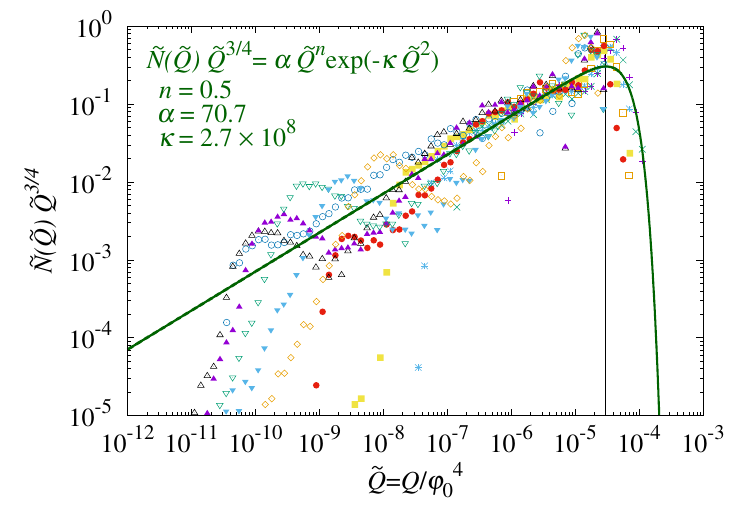} 
\caption{Normalized charge distribution of the Q balls. The thick solid (green) line shows the fitting formula, whose peak is denoted 
in the vertical thin solid (black) line at $\tilde{Q}=3\times 10^{-5}$. 
\label{fig4}}
\end{figure}

\begin{figure}[ht!]
\includegraphics[width=90mm]{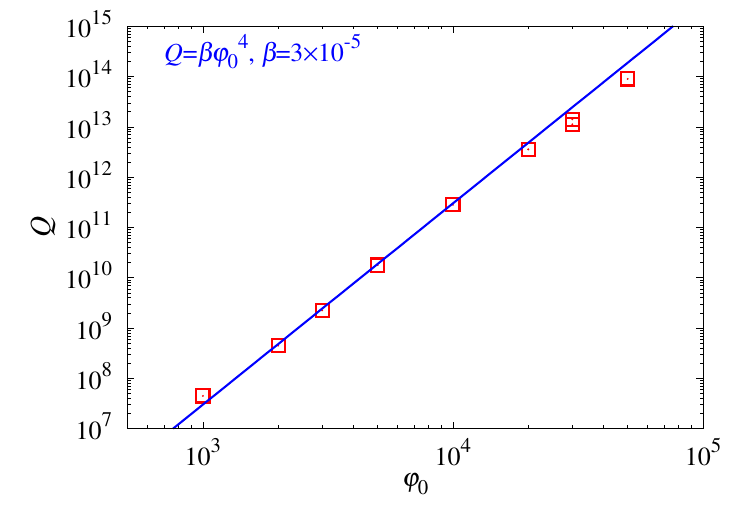} 
\caption{Peak charge of the Q balls. Also shown is the relation (\ref{Qphi2}) in the blue solid line.
\label{fig5}}
\end{figure}

The charge distributions of the Q balls, $\tilde{N}(\tilde{Q})\tilde{Q}^{3/4}$, are displayed in Fig.~\ref{fig4} for all the cases, 
where $\tilde{N}(\tilde{Q})$ is the number of the Q balls in the logarithmic interval of the charge $\tilde{Q}$ and $\tilde{Q}=Q/\varphi_0^4$. 
$Q^{3/4}$ is multiplied because one can locate the peak charge of the Q balls which 
dominate the energy density. See Eq.(\ref{MQ}). We normalize the distribution function so as to give unity when it is 
integrated over the whole charge. We adopt  the fitting formula of the form
\begin{equation}
\tilde{N}(\tilde{Q})\tilde{Q}^{3/4} = \alpha \tilde{Q}^n \exp\left(-\kappa \tilde{Q}^2\right),
\label{Qdist}
\end{equation}
with $n=0.5$ and $\kappa=2.7\times 10^8$, and $\alpha$ is determined as 70.7 from
\begin{equation}
\int \tilde{N}(\tilde{Q})\tilde{Q}^{3/4} d\log\tilde{Q} = 1,
\end{equation}
which is shown in the green solid line in the figure. Notice that those Q balls with smaller charges are not included
for the fit due to low resolutions. At the peak of the distribution, it is equivalent to the relation
\begin{equation}
Q=\beta'\left(\frac{\phi_0}{m}\right)^4 = \beta' \left(\frac{\phi_0}{M_F}\right)^4,
\label{Qphi2}
\end{equation}
with $\beta'=3\times 10^{-5}$. We plot the peak charges of the Q ball in terms of the initial field value 
$\varphi_0$ in Fig.~\ref{fig5}. They all align on the relation (\ref{Qphi2}), shown in the blue solid line in this figure. 

Roughly speaking, the charge of the formed Q ball is determined by the charge of the field inside the horizon at the formation time. 
Since the field starts its oscillation earlier in the logarithmic square potential than in the logarithmic potential at the same initial 
field amplitude, and the fluctuations grow faster, $\beta'$ becomes smaller than $\beta$ in Eq.(\ref{Qphi1}).

Finally, we evaluate the number of the Q balls in the horizon size. As seen in Fig.~\ref{fig2}, the Q-ball formation takes place
at around $a/a_{\rm init}=6$. Therefore, we count the number of the Q balls at that time. Since each simulation has different
lattice spacing, hence the different box size, we need the weighted average the Q-ball numbers with respect to the actual
numbers of each simulation box, which can be estimated as
\begin{equation}
\langle N_{\rm hor} \rangle = \frac{\sum_j N_{{\rm hor},j}N_{{\rm box},j} }{\sum_j N_{{\rm box},j}},
\end{equation}
where $N_{{\rm box},j}$ and $N_{{\rm hor},j}$ are the numbers of the Q balls in the simulation box and the horizon size 
in the $j$-th simulation, respectively. We thus obtain the number of the Q balls as
\begin{equation}
\langle N_{\rm hor} \rangle \simeq 1.4\times 10^4, \ 6.6\times 10^4, \ 1.5\times 10^5,
\end{equation}
with the charge larger than $Q_{\rm peak}$, $0.1 Q_{\rm peak}$ and $0.01 Q_{\rm peak}$, respectively,
where $Q_{\rm peak}$ is the Q-ball charge at the peak of the distribution.

\section{Cosmological and astrophysical consequences of the broad charge distribution}
Let us investigate the differences of cosmological and astrophysical results between the monochromatic and broad distribution 
of the Q-ball charge. \footnote{
We consider the field amplitude smaller than $\phi_{\rm eq}$, which can be as large as $\phi_{\rm max}\simeq 0.64 M_{\rm P}$,
so that the gravity-mediation potential does not affect the dynamics of the field. See Eq.(\ref{Qform}).
}

\subsection{B balls}
We first consider the Q balls with the charge being the baryon number, so-called B balls.
The most striking feature of the B ball is the stability against the decay into nucleons, the lightest particles with the unit baryon number,
if the B-ball charge is large enough. This could be rephrased in terms of the condition on $\omega_Q$ 
as \cite{Kusenko:1997si,Kasuya:2011ix}
\begin{equation}
\omega_Q < b m_N,
\end{equation}
where $m_N$ is the nucleon mass, and $b$ represents the effective baryon number of the fields that constitute the Q ball.
It implies that the Q-ball mass per unit charge should be smaller than the nucleon mass to avoid the Q-ball decay into
nucleons. Therefore, Q balls with a charge larger than $Q_{\rm cr}$ can be the dark matter of the universe, where
\begin{equation}
Q_{\rm cr} \simeq 2.5\times 10^{31} \left(\frac{b}{1/3}\right)^{-4} \left(\frac{M_F}{10^6~{\rm GeV}}\right)^4.
\label{Qcr}
\end{equation}

\begin{figure}[ht!]
\includegraphics[width=85mm]{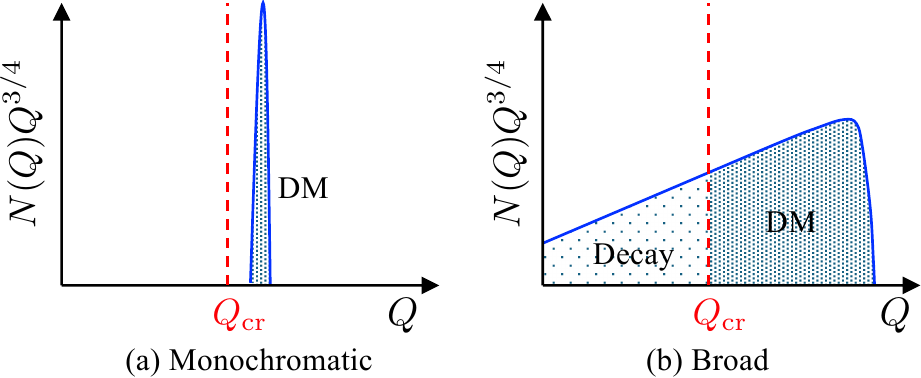} 
\caption{Sketches of (a) monochromatic and (b) broad distribution of the B balls.
\label{fig6}}
\end{figure}

As for the dark matter Q balls, there is little difference between the monochromatic and broad distributions.
One only needs to integrate the charge distribution above $Q_{\rm cr}$ to calculate
the dark matter Q-ball abundance as can be seen in Fig.~\ref{fig6}. From Eq.(\ref{Qdist}), we have still about 1/4 (3/4) of the formed
Q balls to be the dark matter even if $Q_{\rm cr} = Q_{\rm peak} (0.1Q_{\rm peak})$ for the broad distribution.

However, one must consider the effects of the decay of those Q balls with smaller charges than $Q_{\rm cr}$, as seen in the right
panel in Fig.~\ref{fig6}. Let us first investigate the situation that would explain the baryon number of the universe simultaneously 
by the single flat direction through the decayed Q balls. 

In the monochromatic distribution case, enough baryon numbers cannot be provided, where they are evaporated from the 
surface of the formed Q balls \cite{Kasuya:2014ofa}. One may need two flat directions: one contributes to form the dark 
matter Q balls, and the other produces the baryon numbers by the decay of the unstable Q balls \cite{Kasuya:2014ofa}.

On the other hand, both stable and unstable Q balls coexist in the broad distribution for the single flat direction, 
where the former constitutes the dark matter, while the latter may explain the baryon number of the universe. 
Since the right amount of baryon numbers should exist before the big bang nucleosynthesis (BBN), the simplest scenario
is that the unstable Q balls decay before BBN.

Q-ball decay occurs if some decay particles carry the same kind of the charge of the Q ball and the mass of all  the 
decay particles is less than the mass of the Q ball per unit charge $\omega_Q$. Since the decay products are fermions, 
once the Fermi sea is filled, further decay proceeds only when produced fermions
escape from the surface of the Q ball. The upper bound of the decay rate is thus determined by the maximum 
outgoing flow of the fermions \cite{Cohen:1986ct}. This saturation takes place when the field value is large inside the
Q ball, which is the case here. One can thus estimate the decay rate as \cite{Cohen:1986ct,Kawasaki:2012gk,Kamada:2012bk}
\begin{equation}
\Gamma_Q \simeq \frac{1}{Q}\frac{\omega_Q^3}{12\pi^2}4\pi R_Q^2 
\simeq \frac{\sqrt{2}\pi^2\zeta}{3} M_F Q^{-5/4},
\label{decayrate}
\end{equation}
where we use Eqs.(\ref{RQ}) and (\ref{omegaQ}) in the last equality.
Then the Q-ball charge should be less than
\begin{equation}
Q_{\rm D} \simeq 1.7 \times 10^{25} \left(\frac{M_F}{10^6~{\rm GeV}}\right)^{4/5},
\label{Qdecay}
\end{equation}
where we set $\Gamma_Q^{-1} \lesssim 1$~s. 

\begin{figure}[ht!]
\includegraphics[width=85mm]{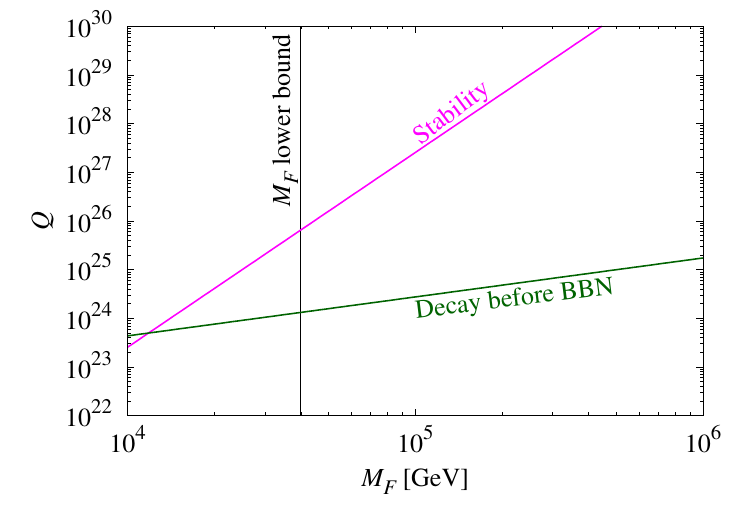} 
\caption{Charges of the Q balls which are stable against the decay into nucleons (\ref{Qcr}) and which decay before BBN (\ref{Qdecay})
are shown in magenta and green lines, respectively.
\label{fig7}}
\end{figure}

Then, from Eqs.(\ref{Qcr}) and (\ref{Qdecay}), the requirement that the unstable Q balls decay before BBN leads to
$M_F < 1.2 \times 10^4$~GeV, which is smaller than the lower bound of $M_F$ in Eq.(\ref{MF}).
Therefore, the simplest scenario cannot be realized, and the Q balls within some range of the charge decay after BBN 
for allowed $M_F$, as seen in Fig.~\ref{fig7}.

The abundance of the decay particles after BBN is strictly constrained for successful BBN, which can be written 
as \cite{Kawasaki:2017bqm}
\begin{equation}
\frac{\rho_Q^{\rm (decay)}}{s} \lesssim 10^{-14}~{\rm GeV},
\label{BBN}
\end{equation}
where $\rho_Q^{\rm (decay)}$ is the energy density of the Q balls with $Q<Q_{\rm cr}$, and $s$ is the entropy density. 
We can thus estimate the upper limit of the ratio of the Q balls that decay after BBN and the dark matter Q balls as
\begin{equation}
\frac{\rho_Q^{\rm (decay)}}{\rho_{\rm DM}} =
\frac{\rho_Q^{\rm (decay)}}{s} \left(\Omega_{\rm DM} \frac{\rho_{\rm cr,0}}{s_0}\right)^{-1}
 \lesssim  2.3\times 10^{-5}.
 \label{cond_decay}
\end{equation}
In the last inequality we use the upper bound (\ref{BBN}), $\Omega_{\rm DM}h^2=0.120$ \cite{Planck:2018vyg}, and
$\rho_{\rm cr,0}/s_0 = 3.63\times 10^{-9}h^2$~GeV, where $h$ is the Hubble constant in units of 100~km/s/Mpc.
Therefore, the unstable Q balls cannot provide enough baryon numbers of the universe, since the observed baryon to 
dark matter ratio is estimated as $0.186$ \cite{Planck:2018vyg}.

Now we must check that the condition of the amount of the unstable Q balls (\ref{cond_decay}) could be satisfied so that
the stable B balls would be the dark matter of the universe. Since the unstable-to-stable Q-ball ratio is calculated as
\begin{equation}
\frac{\rho_Q^{\rm (decay)}}{\rho_Q^{\rm DM}}
=\frac{\displaystyle{\int_0^{\tilde{Q}_{\rm cr}} N(\tilde{Q})\tilde{Q}^{3/4}\frac{d\tilde{Q}}{\tilde{Q}}}}
{\displaystyle{\int_{\tilde{Q}_{\rm cr}}^{\infty} N(\tilde{Q})\tilde{Q}^{3/4}\frac{d\tilde{Q}}{\tilde{Q}}}},
\end{equation}
we find that the condition (\ref{cond_decay}) holds for $\tilde{Q}_{\rm cr} \lesssim 2.6\times 10^{-14}$, which results in
\begin{equation}
Q_{\rm cr} \lesssim 3.0 \times 10^{-14} \left(\frac{\phi_0}{M_F}\right)^4.
\end{equation}
From Eq.(\ref{Qcr}), we arrive at the upper limit on $M_F$ as
\begin{equation}
M_F \lesssim 3.0\times 10^6 \left(\frac{\phi_0}{\phi_{\rm max}}\right)^{1/2} \left(\frac{b}{1/3}\right)^{1/2}.
\label{MFBBN}
\end{equation}

\begin{figure}[ht!]
\includegraphics[width=85mm]{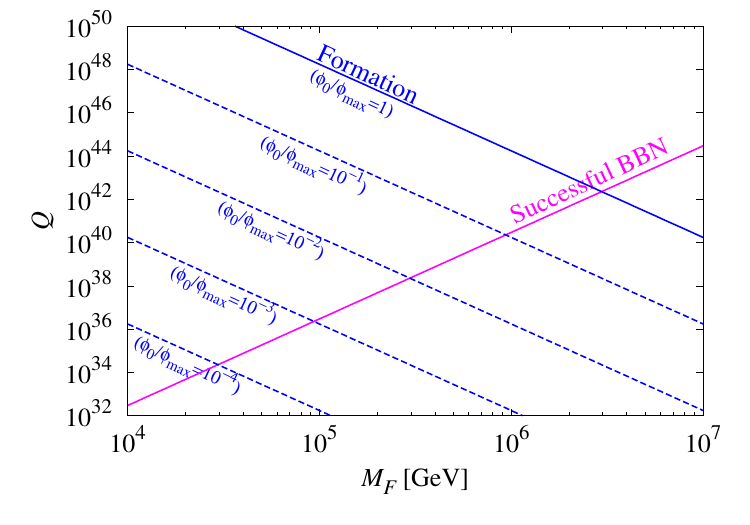} 
\caption{Charge of the Q ball at the formation (\ref{Qphi2}) and that of the decayed Q balls that do not ruin the successful 
BBN (\ref{Qdecay}) are shown in blue and magenta lines, respectively.
\label{fig8}}
\end{figure}

The condition (\ref{cond_decay}) can be understood in another way: Q balls should be arranged to form such that $Q_{\rm cr}$ is  
sufficiently small. Thus it can be rephrased as
\begin{equation}
Q_*\equiv Q_{\rm cr} \frac{\tilde{Q}_{\rm peak}}{\tilde{Q}_{\rm cr}}
=2.9\times 10^{40}\left(\frac{b}{1/3}\right)^{-4}\left(\frac{M_F}{10^6~{\rm GeV}}\right)^4,
\label{Qcrpeak}
\end{equation}
should be smaller than $Q_{\rm form}$ (\ref{Qphi2}). Here, $\tilde{Q}_{\rm peak} =\beta'=3\times 10^{-5}$ and Eq.(\ref{Qcr})
are used. We plot Eqs.(\ref{Qphi2}) and (\ref{Qcrpeak}) in blue and magenta lines, respectively, in Fig.~\ref{fig8}. One can perceive that
 Eq.(\ref{MFBBN}) is seen there.

Anyway, B balls can therefore be the dark matter of the universe for a broad range of $M_F$, such as 
$4\times 10^4~{\rm GeV} \lesssim M_F \lesssim 3.0\times 10^6$~GeV.

\subsection{L balls}
An L ball is a Q ball whose charge is the lepton number. 
Here we reconsider the scenario that L balls decay into positrons at present which may explain the
511 keV gamma-ray flux from the galactic center or avoid its overflux \cite{Kasuya:2005ay,Kasuya:2024ldq} with the broad
charge distribution of formed Q balls, and seek the possibility that L balls with larger charges in the Q-ball distribution
could be the dark matter of the universe as well.

\begin{figure}[ht!]
\includegraphics[width=85mm]{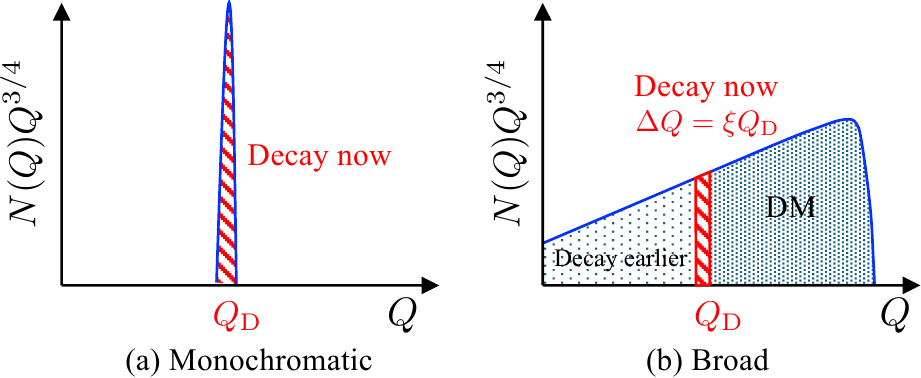} 
\caption{Sketches of monochromatic and broad distribution of the L balls.
\label{fig9}}
\end{figure}

For those Q balls that decay at present, 
their lifetime is set to be $\tau_Q = \Gamma_Q^{-1} \simeq t_0 \simeq 13.8$~Gyr, which leads to 
the charge of the Q ball as
\begin{equation}
Q_{\rm D}  \simeq 1.7\times 10^{38} \left(\frac{M_F}{4\times 10^4~{\rm GeV}}\right)^{4/5}.
\label{decay}
\end{equation}
This charge is smaller than the maximum charge of the formed gauge-mediation type Q balls for small enough $M_F$, 
estimated as
\begin{equation}
Q_{\rm form} \le \beta' \left(\frac{\phi_{\rm eq}}{M_F}\right)^4 \simeq 6.8\times 10^{49}\left(\frac{M_F}{4\times 10^4~{\rm GeV}}\right)^{-4},
\label{Qform}
\end{equation}
where $\phi_{\rm eq} = \phi_{\rm max} \simeq 0.64M_{\rm P}$, which comes from the upper bound of $M_F$
in Eq	.(\ref{MF}), is used in the last equality.
Therefore, Q balls with a larger charge than $Q_{\rm D}$ have a longer lifetime to become the dark matter, while the smaller
Q balls have decayed earlier. See Fig.~\ref{fig9}.

Below, we specify the flat direction which constitutes the L ball as $L_iL_je_k$ ($i\ne j$), where $L$ and $e$ denote
SU(2)$_L$ doublet and singlet sleptons, respectively. 

\subsubsection{$\tilde{\nu}_\mu\tilde{e}^-\tilde{e}^+$ direction with $\phi_0=\phi_{\rm max}$}

Let us consider the simplest example where $L_iL_je_k=\tilde{\nu}_\mu\tilde{e}^-\tilde{e}^+$. In this case, the L balls decay into 
$\nu_\mu \nu_e \bar{\nu}_e$ in the first place. As the charge decreases, the decay channel into $e^-$ and $e^+$ opens when 
$\omega_Q$ becomes larger than the electron (positron) mass $m_e$. This happens when
\begin{equation}
Q<Q_{e^+} \equiv \left(\frac{\sqrt{2}\pi\zeta M_F}{m_e}\right)^4 = 9.1\times 10^{36} \left(\frac{M_F}{4\times 10^4~{\rm GeV}}\right)^4.
\label{Qep1}
\end{equation}
Then, the L balls can additionally decay into $\nu_\mu e^-e^+$. The created positrons may annihilate with surrounding electrons
to produce the 511 keV gamma rays at the galactic center \cite{Kasuya:2005ay,Kasuya:2024ldq}.

The ratio of the density of the decay products and the Q-ball dark matter is evaluated by
\begin{equation}
\frac{\Omega_{\rm dec}}{\Omega_{\rm DM}}
=\frac{\displaystyle{\int_{\tilde{Q}_{\rm dec}}^{(1+\xi)\tilde{Q}_{\rm dec}} N(\tilde{Q})\tilde{Q}^{3/4}\frac{d\tilde{Q}}{\tilde{Q}}}}
{\displaystyle{\int_{(1+\xi)\tilde{Q}_{\rm dec}}^{\infty} N(\tilde{Q})\tilde{Q}^{3/4}\frac{d\tilde{Q}}{\tilde{Q}}}}
\simeq 1.3 \times 10^{-8},
\end{equation}
where $\tilde{Q}_{\rm dec}=3\times 10^{-5} (Q_{\rm D}/Q_{\rm form})\simeq 7.5\times 10^{-17}$. Here we adopt
$\phi_0=\phi_{\rm max}$ and set $\xi=(4/5)(Q_{e^+}/Q_{\rm D})^{5/4} \simeq 0.02$, where we assume that Q balls 
which decay at present have charges with width $\Delta Q_{\rm D} =\xi Q_{\rm D}$. See Appendix B.

\begin{figure}[ht!]
\includegraphics[width=85mm]{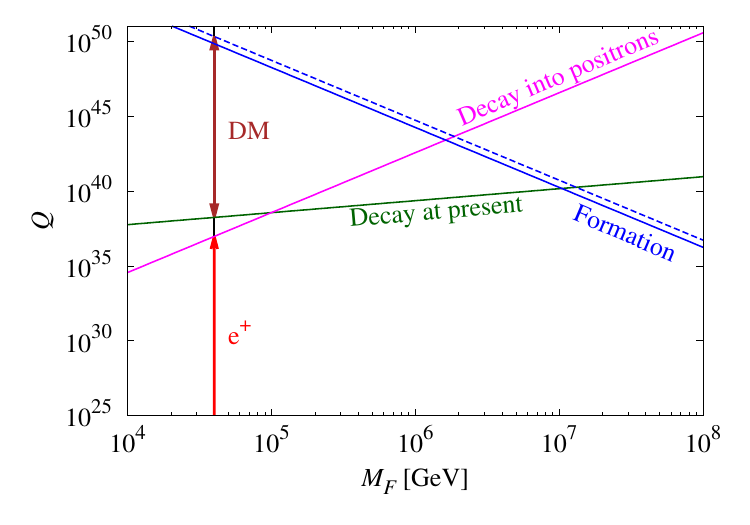} 
\caption{Charges of the Q balls which decay at present (\ref{decay}) and when the decay channel into positrons 
opens (\ref{Qep1}) are shown in green and magenta lines, respectively. Blue solid and dashed lines respectively represent
the peak and maximum charges of the formed Q balls.
\label{fig10}}
\end{figure}

Since the charge fraction of the Q ball which decay into positrons over the decaying Q balls is obtained from 
Eqs.(\ref{decay}) and (\ref{Qep1}) as
\begin{equation}
\frac{Q_{e^+}}{Q_{\rm D}} \simeq 5.4\times 10^{-2} \left(\frac{M_F}{4\times 10^4~{\rm GeV}}\right)^{16/5},
\end{equation}
then the ratio of the density parameters is derived as
\begin{equation}
\frac{\Omega_{e^+}}{\Omega_{\rm D}} = \left(\frac{Q_{e^+}}{Q_{\rm D}} \right)^{3/4} 
\simeq 0.11 \left(\frac{M_F}{4\times 10^4~{\rm GeV}}\right)^{12/5}.
\end{equation}
Therefore, the ratio of the density parameters of the positrons from the Q-ball decay and
the Q-ball dark matter is given by
\begin{equation}
\frac{\Omega_{e^+}}{\Omega_{\rm DM}}=\frac{\Omega_{e^+}}{\Omega_{\rm D}}\frac{\Omega_{\rm D}}{\Omega_{\rm DM}}
\simeq 1.4 \times 10^{-9} \left(\frac{M_F}{4\times 10^4~{\rm GeV}}\right)^{12/5},
\end{equation}
which is larger than the upper bound $2.7\times 10^{-10}$, derived in the Appendix A.
This fraction is indicated by `DM' and ``$e^+$'' in Fig.~\ref{fig10}.
Since $M_F$ cannot be smaller than $4\times 10^4$~GeV, the simple scenario of $\tilde{\nu}_\mu\tilde{e}^-\tilde{e}^+$ direction 
does not work.

\subsubsection{$\tilde{\nu}_\tau\tilde{\mu}^-\tilde{e}^+$ direction with $\phi_0=\phi_{\rm max}$}

\begin{figure}[ht!]
\includegraphics[width=85mm]{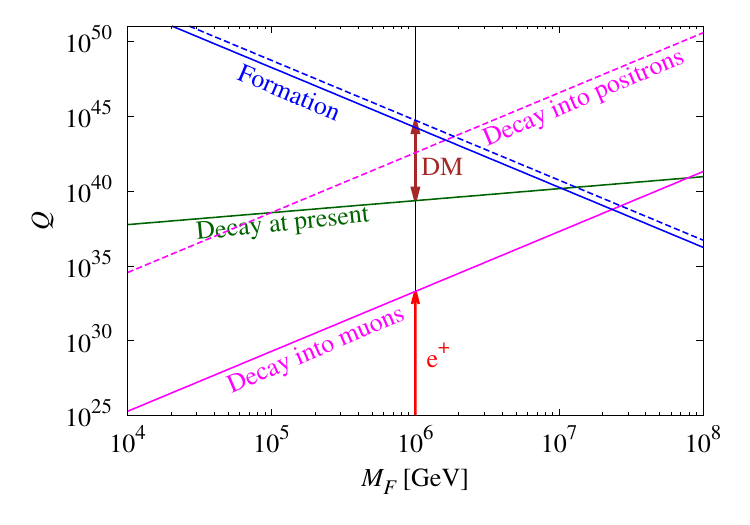} 
\caption{Same as in Fig.~\ref{fig10}. Also shown in the dashed magenta line is the Q-ball charge when the decay channel
into muons opens (\ref{Qmu}).
\label{fig11}}
\end{figure}

Now we move on to the case of $L_iL_je_k=\tilde{\nu}_\tau\tilde{\mu}^-\tilde{e}^+$. The Q balls decay into
$\nu_\tau\nu_\mu\bar{\nu}_e$ in the first place. However, positrons cannot be produced until the decay channel into
muons opens, since the decay products would be $\nu_\tau\mu^-e^+$. In this case, the charge fraction that creates
positrons will be smaller, which leads to the fact that the scenario may work for larger $M_F$, as shown in Fig.~\ref{fig11}.

In this figure, we also plot the charge of the Q balls when the decay into muons are allowed in the solid magenta line.
It is obtained from $\omega_Q > m_\mu$, where $m_\mu$ is the muon mass, as
\begin{equation}
Q< Q_\mu \equiv 2.0 \times 10^{33} \left(\frac{M_F}{10^6~{\rm GeV}}\right)^4,
\label{Qmu}
\end{equation}
where we choose $M_F=10^6$~GeV; also the charges at the formation and that of the decay at present
are estimated as
\begin{eqnarray}
Q_{\rm form} & \simeq & 1.7\times 10^{44} \left(\frac{M_F}{10^6~{\rm GeV}}\right)^{-4}, \\
Q_{\rm D} & \simeq & 2.2 \times 10^{39} \left(\frac{M_F}{10^6~{\rm GeV}}\right)^{4/5},
\label{decay2}
\end{eqnarray}
which leads to $\tilde{Q}_{\rm dec}\simeq 3.9\times 10^{-10}$. Here we assume $\phi_0=\phi_{\rm max}$.

We can follow a similar argument as in the previous case. 
The ratio of the density of the decay products and the Q-ball dark matter is evaluated by
\begin{equation}
\frac{\Omega_{\rm dec}}{\Omega_{\rm DM}}
=\frac{\displaystyle{\int_{\tilde{Q}_{\rm dec}}^{(1+\xi)\tilde{Q}_{\rm dec}} N(\tilde{Q})\tilde{Q}^{3/4}\frac{d\tilde{Q}}{\tilde{Q}}}}
{\displaystyle{\int_{(1+\xi)\tilde{Q}_{\rm dec}}^{\infty} N(\tilde{Q})\tilde{Q}^{3/4}\frac{d\tilde{Q}}{\tilde{Q}}}}
\simeq 3.0 \times 10^{-11},
\end{equation}
for $\xi=2\times 10^{-8}$. Since the positron production takes place after the decay channel into muons opens, 
the fraction of the density parameters of the Q balls that decay into positrons is obtained from 
Eqs.(\ref{Qmu}) and (\ref{decay2}) as
\begin{equation}
\frac{\Omega_{e^+}}{\Omega_{\rm D}} = \left(\frac{Q_\mu}{Q_{\rm D}} \right)^{3/4} 
\simeq 2.9\times 10^{-5} \left(\frac{M_F}{10^6~{\rm GeV}}\right)^{12/5}.
\end{equation}
Therefore, the ratio of the density parameters of the positrons from the Q-ball decay and
the Q-ball dark matter is given by
\begin{equation}
\frac{\Omega_{e^+}}{\Omega_{\rm DM}}=\frac{\Omega_{e^+}}{\Omega_{\rm D}}\frac{\Omega_{\rm D}}{\Omega_{\rm DM}}
\simeq 8.5\times 10^{-16} \left(\frac{M_F}{10^6~{\rm GeV}}\right)^{12/5},
\label{abund1}
\end{equation}
which is well below the upper bound derived as $1.8\times 10^{-13}$ in the Appendix A. Thus we cannot explain the
observed 511 keV gamma flux by the decay of the smaller L balls. 

Notice that there is more stringent
constraints from X-ray observations by inverse Compton scattering \cite{Cirelli:2023tnx}. We adopt the conservative bound
on the lifetime as $\tau_{\rm ddm}\simeq 3\times 10^{24}$~s in the case of the decaying dark matter. It can be rephrased as
\begin{equation}
\frac{\Omega_{e^+}}{\Omega_{\rm DM}} \lesssim \frac{\xi t_0}{\tau_{\rm ddm}} 
\simeq 3\times 10^{-15} \left(\frac{\xi}{2\times 10^{-8}}\right),
\end{equation}
in our case, where the derived abundance (\ref{abund1}) is safely below this constraint. Therefore, large L balls
can be the dark matter of the universe for $M_F \lesssim 10^6$~GeV.

\subsubsection{$\tilde{\nu}_\tau\tilde{\mu}^-\tilde{e}^+$ direction with smaller $\phi_0$}

\begin{figure}[ht!]
\includegraphics[width=85mm]{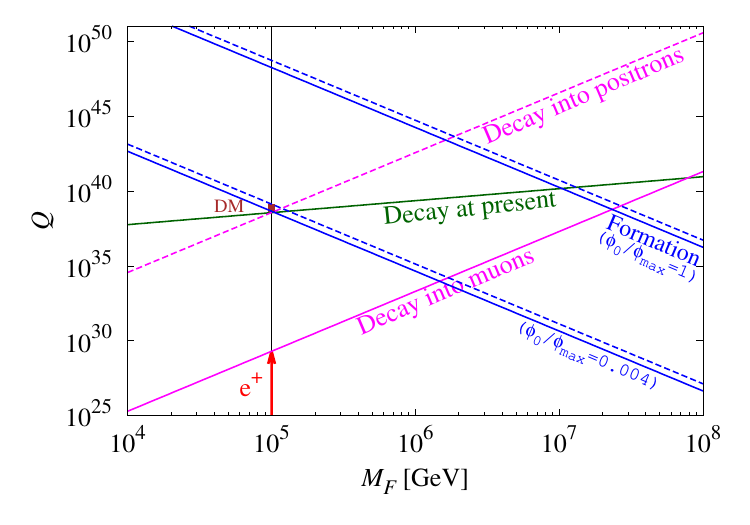} 
\caption{Same as in Fig.~\ref{fig10}. Also shown is the charge of the formed Q balls for a smaller initial amplitude of 
$\phi_0/\phi_{\rm max}=0.004$.
\label{fig12}}
\end{figure}

So far we set $\phi_0 =\phi_{\rm max}$. Now let us investigate the case of a smaller initial amplitude of the field. This is realized
for smaller $M_F$, as can be seen in Fig.~\ref{fig12}. A smaller positron fraction is compensated by a smaller fraction of the dark
matter Q balls. 

The Q-ball charges at the formation, which decay at present, and when the decay into muons is allowed
are respectively estimated as
\begin{eqnarray}
Q_{\rm form} & \simeq & 4.4\times 10^{38} \left(\frac{M_F}{10^5~{\rm GeV}}\right)^{-4}, \\
\label{decay3}
Q_{\rm D} & \simeq & 3.5\times 10^{38}  \left(\frac{M_F}{10^5~{\rm GeV}}\right)^{4/5},\\
Q_\mu & \simeq & 2.0 \times 10^{29} \left(\frac{M_F}{10^5~{\rm GeV}}\right)^4,
\label{Qmu2}
\end{eqnarray}
for $\phi_0=0.004\phi_{\rm max}$, which leads to $\tilde{Q}_{\rm dec} \simeq 2.4\times 10^{-5}$.

Then the ratio of the density of the decay products and the Q-ball dark matter is calculated as
\begin{equation}
\frac{\Omega_{\rm dec}}{\Omega_{\rm DM}}
=\frac{\displaystyle{\int_{\tilde{Q}_{\rm dec}}^{(1+\xi)\tilde{Q}_{\rm dec}} N(\tilde{Q})\tilde{Q}^{3/4}\frac{d\tilde{Q}}{\tilde{Q}}}}
{\displaystyle{\int_{(1+\xi)\tilde{Q}_{\rm dec}}^{\infty} N(\tilde{Q})\tilde{Q}^{3/4}\frac{d\tilde{Q}}{\tilde{Q}}}}
\simeq 1.9 \times 10^{-12},
\end{equation}
for $\xi=2\times 10^{-12}$ and $\phi_0/\phi_{\rm max}=0.004$. On the other hand,
the fraction of the density parameters of the Q balls that decay into positrons is obtained from 
Eqs.(\ref{decay3}) and (\ref{Qmu2}) as
\begin{equation}
\frac{\Omega_{e^+}}{\Omega_{\rm D}} = \left(\frac{Q_\mu}{Q_{\rm D}} \right)^{3/4} 
\simeq 1.1\times 10^{-7} \left(\frac{M_F}{10^5~{\rm GeV}}\right)^{12/5}.
\end{equation}
Therefore, the ratio of the density parameters of the positrons from the Q-ball decay and
the Q-ball dark matter is estimated as
\begin{equation}
\frac{\Omega_{e^+}}{\Omega_{\rm DM}}=\frac{\Omega_{e^+}}{\Omega_{\rm D}}\frac{\Omega_{\rm D}}{\Omega_{\rm DM}}
\simeq 2.2\times 10^{-19} \left(\frac{M_F}{10^5~{\rm GeV}}\right)^{12/5},
\label{abund2}
\end{equation}
which is well below the upper bound derived as $1.9\times 10^{-17}$ in the Appendix A. Thus we cannot explain the
observed 511 keV gamma flux by the decay of the smaller L balls also in this case.

Again the more stringent constraint comes from X-ray observations by inverse Compton scattering \cite{Cirelli:2023tnx}.
In this case it reads as $\Omega_{e^+}/\Omega_{\rm DM}\lesssim 3\times 10^{-19}$ for $\xi\simeq 2\times 10^{-12}$,
which is safely above the derived abundance (\ref{abund2}). Then the large L balls can be the dark matter for 
$\phi_0/\phi_{\rm max}\gtrsim 0.004$ for $M_F=10^5$~GeV.

\section{Conclusion}
We have numerically investigated the formation of the gauge-mediation type Q balls in the logarithmic square
potential on the $1000^3$ three-dimensional lattices, and obtained the broad charge distribution for this type
of the  Q balls for the first time, to the best of our knowledge. The charge of the Q ball at the peak of the distribution is a bit smaller than
what we had obtained as the average of the largest tens of the Q balls in the logarithmic potential. 

We have discussed some impacts of the broad distribution on cosmology and astrophysics. The broad charge
distribution may imply that both stable and unstable B balls exist. We have found that the latter would not ruin the 
successful BBN, but we cannot explain simultaneously the dark matter and the baryon number of the universe
by a single flat direction. On the other hand, 
L balls with larger charges have longer lifetime than the present age of the universe to be the dark matter of the universe,
while the smaller Q balls decay into light charged leptons, which may contribute to the X and/or gamma rays. We have found
that the flux could be well below the observational constraints.

\section*{Acknowledgments} 
This work is supported by JSPS KAKENHI Grants No. 20H05851(M. K.) and No. 21K03567(M. K.).

\section*{DATA AVAILABILITY} 
The data that support the findings of this article are not publicly available upon publication because it is not technically 
feasible and/or the cost of preparing, depositing, and hosting the data would be prohibitive within the terms of this 
research project. The data are available from the authors upon reasonable request.

\appendix

\section{Rough constraint on $\Omega_{e^+}/\Omega_{\rm DM}$}
Q-ball decay  creates positrons with energy $\lesssim \omega_Q$ which may annihilate with electrons
at the Galactic Center to produce 511~keV gamma rays, and simultaneously remaining Q balls explain the dark matter 
of the universe. Since the morphology of the dark matter halo 
in our galaxy is still not known, we make a very rough estimate for the upper bound of the fraction of the Q balls that
decay into positrons, according to Refs.~\cite{Hooper:2004qf,Kasuya:2024ldq}. Assuming the half of the total 511~keV flux is 
emitted from an angular region of $9^\circ$ circle, we have
\begin{equation}
\frac{M_{(<9^\circ)}}{\omega_Q \Delta\tau_Q} \frac{\Omega_{e^+}}{\Omega_{\rm DM}} f_{e^+} \left[\frac{f}{4}+(1-f)\right]
=\frac{1}{2} \Phi_{511} 4\pi R_{\rm GC}^2,
\end{equation}
where $f$ is the fraction of positrons which annihilates via positronium, $\Phi_{511}(\simeq 10^{-3}$~cm$^{-2}$s$^{-1}$)
is the observed total flux of 511~keV line, $R_{\rm GC}\simeq 8.23$~kpc, and $\Delta\tau_Q=\xi'\tau_Q$, where
$\xi' = (5/4)\xi$. See Eq.(\ref{xiprime}).
We set $\tau_Q \simeq t_0$. The total mass within the $9^\circ$ circle is given by
\begin{equation}
M_{(<9^\circ)} = \int_{(<9^\circ)} \rho_{\rm DM}(r) 4\pi r^2 dr,
\end{equation}
where we assume $\rho_{\rm DM}$ as the Navarro-Frenk-White profile \cite{Navarro:1995iw}.
Therefore, we obtain the constraint on the ratio of the density parameters as
\begin{equation}
\frac{\Omega_{e^+}}{\Omega_{\rm DM}}  \lesssim 8.8\times 10^{-10} 
\left(\frac{\xi}{0.02}\right)\left(\frac{f_{e^+}}{1/6}\right) 
\left(\frac{\omega_Q}{m_{e^+}}\right),
\label{511}
\end{equation}
where we choose the $LLe=\tilde{\nu}_\mu\tilde{e}^-\tilde{e}^+$ direction. In the $LLe=\tilde{\nu}_\tau\tilde{\mu}^-\tilde{e}^+$ 
direction case, since positrons are produced just after $\omega_Q$ becomes larger than the muon mass, $\omega_Q=m_\mu$,
so that $\Omega_{e^+}/\Omega_{\rm DM} \lesssim 1.9\times 10^{-13}$ for $M_F=10^6$~GeV, 
while $\Omega_{e^+}/\Omega_{\rm DM} \lesssim 1.9\times 10^{-17}$ for $M_F=10^5$~GeV.

\section{Value of $\xi$}
Here we estimate the range of the charge of the Q balls which decay into positrons at present time. Since the charge decreasing
rate of the Q ball is given by
\begin{equation}
-\frac{dQ}{dt}={\cal A} Q^{-1/4},
\end{equation}
where ${\cal A}=\sqrt{2}\pi^2 \zeta M_F/3$ is a constant depending on $M_F$ [see Eq.(\ref{decayrate})], then the evolution of the
charge reads as \cite{Kasuya:2022cko}
\begin{equation}
Q(t)=Q_i \left( 1-\frac{t}{\tau(Q_i)}\right)^{4/5},
\end{equation}
where $Q_i$ is the initial charge of the Q ball and $\tau(Q)=4Q^{5/4}/5{\cal A}$ is the lifetime of the Q ball with the charge $Q$. 

The decaying Q balls which contribute to the production of positrons at present must satisfy the condition 
$\tau(Q_{\rm D}) -\tau(Q_x) \lesssim t_0 \lesssim \tau(Q_{\rm D})$ with $x=e^+$ or $\mu$. This implies that the span of the time is
$\Delta t=\tau(Q_x)$. Since
\begin{equation}
\frac{d\tau(Q)}{dQ}=\frac{Q^{1/4}}{\cal A}=\frac{5}{4} \frac{\tau(Q)}{Q},
\end{equation}
we obtain
\begin{equation}
\frac{\Delta t}{\tau(Q)}=\frac{5}{4}\frac{\Delta Q}{Q}.
\end{equation}
We can therefore estimate $\xi$ as
\begin{equation}
\xi = \frac{\Delta Q}{Q_{\rm D}}=\frac{4}{5} \frac{\tau(Q_x)}{\tau(Q_{\rm D})}=\frac{4}{5}\left(\frac{Q_x}{Q_{\rm D}}\right)^{5/4}.
\label{xiprime}
\end{equation}



\end{document}